\documentclass[12pt]{article}
\usepackage{sc3conf}
\usepackage{amsfonts}
\usepackage{graphicx}
\usepackage{epsfig} 

\begin{document}


\title{Search for distortions in the spectrum of the Cosmic Microwave Radiation}


\authors{G.~Sironi,\adref{1} E.~Battistelli,\adref{2} G.~Boella,\adref{1}
F.~Cavaliere,\adref{3} M.~Gervasi,\adref{1} A.~Passerini,\adref{1} A.~Tartari,\adref{1} and
  M.~Zannoni\adref{4}}


\addresses{\1ad Physics Department G.~Occhialini, University of Milano-Bicocca, Milano Italy
  \nextaddress
  \2ad Physics Department, University of Roma La Sapienza, Rome Italy,
  \nextaddress
  \3ad Physics Department, University of Milano, Milano Italy
  \nextaddress
  \4ad IASF/CNR - Milano Italy}


\maketitle


\begin{abstract}
  We present preliminary results of TRIS, an experiment dedicated to the
  search of deviations from a pure planckian distribution in the spectrum of
  the Cosmic Microwave Background at frequencies close to 1 GHz.
\end{abstract}

\section{Introduction}
According the Standard Model the expected spectrum of the Cosmic Microwave
Background (CMB) is an equilibrium, (blackbody), spectrum therefore a plot of
measured values of the CMB thermodinamic temperature versus frequency should
be perfectly flat. At the second order level we expect however deviations
from equilibrium ({\it distortions}) because a variety of processes which
accompanied the evolution of the Universe (e.g. turbulence dissipation,
particle annihilation and so on) may have disturbed it injecting energy
which, through matter - radiation interactions, modify the photon
distribution. Subsequent interactions gradually redistribute the photons and,
under proper conditions, completely cancel the distortions (see for instance
\cite{part}). The entire process, described by the Kompaneets Equation
{\cite{komp}, has been studied by many authors (e.g. \cite{dane,buri,daly})
What remains and we see today depends on the epoch when the energy was
injected. Marking the epoch of the energy injection by the red shift
parameter $Z$ three scenarios are possible:

i)$Z\geq 10^7$: at that epoch the velocity of the matter radiation
interactions was so high that the expanding Universe was a sequence of
equilibrium states. The energy $\Delta E$ injected at that epoch  was
therefore redistributed over the entire spectrum and the Planck distribution
reestablished almost immediately. No distortion is therefore expected today;

ii)$10^7 \geq Z \geq 10^4$: the reaction velocity and the time elapsed since
the energy injections are insufficient to reestablish  equilibrium but enough
to produce a Bose Einstein semiequilibrium condition, charachterized by the
{\it chemical potential} $\mu$, proportional to the fraction $\Delta E/E$ of
energy injected, and by the Comptonization parameter $y$. Therefore if energy
injections occurred at $10^7 \geq Z \geq 10^4$, today we should see
distortions, among which particularly interesting are distortions at
frequencies close to 1 GHz: a deep in the distribution of $T_{CMB}$ versus
$\nu$ (see for instance \cite{dane, buri}). The precise frequency of the deep
depends on the barion density $\Omega_b$.

iii)$Z \leq 10^4$: after recombination the matter radiation-interactions were
drastically reduced, therefore distortions produced at $Z \leq 10^4$ remain
and keep track of the process which generated them.

The effective scenario one observes can be very complicated being the
combination of a variety of processes which injected energy at different
epochs. It is however worth study it and try to detect distortions, unique
probes of the conditions existing in the Universe up to $Z\simeq 10^6 - 10^7$,
well before the condensations we see today began to form.

\section{Measurements of the CMB spectrum}

The most accurate measurement of the CMB spectrum was made ten years ago
\cite{Math1, Math2} with FIRAS,  an absolute radiometer on board the COBE
satellite. It shows that between 60 and 600 GHz an extremely good description
of the CMB spectrum is given by the spectrum of a blackbody with a
temperature $T_{CMB} = (2.725 \pm 0.002 )K$. Deviations from the plankian
shape, if present, are $\leq 0.005 \%$.
  Outside the 60 - 600 GHz frequency range explored by FIRAS the measured values of $T_{CMB}$ are less
accurate \footnote{It has been suggested recently (\cite{batt}) that the
blackbody reference used by FIRAS to calibrate its system had deviations from
the distribution the assumed by the observers. Even if this fact will be
confirmed the FIRAS results will remain the most accurate in literature}. Here
in fact we are far from the maximum of the CMB brightness distribution,
therefore the importance of the foregrounds (galactic synchrotron and blend
of unresolved extragalactic sources on the low frequency side, dust emission
at high frequencies) is higher. Moreover all the observations were made in
worse conditions than in space, on a dedicated satellite.

Above 600 GHz observations were extended with a rocket borne radiometer which
set an upper limit of 3 K to $T_{CMB}$ at 900 GHz \cite{gush}.

Below 60 GHz all the observations were made with radiometers installed at
ground level or on stratospheric balloons, therefore data are comtaminated by
the emission of the earth atmospheric layer above the radiometer and by the
fraction of the environment radiation the antennae pick up through side and
back lobes. In this frequency range the 2.5 - 90 GHz interval was studied 20
years ago by the White Mt. collaboration, using absolute radiometers with
geometrically scaled antennae \cite{whit}. The White Mt results, combined
with independent observations at balloon altitudes \cite{wilk}, set a $1\%$
upper limit to deviations from the Planck distribution between $\simeq 10
GHz$ and $60 GHz$ and a $10\%$ upper limit between 3 and 10 GHz.

 Below $5 GHz$ in literature there are more than 20 independent measurements of
$T_{CMB}$, made in about 40 years. They are listed in Table 1. We quote the
error bars reported by the observers, it is however worth to note that in some
cases these errors probably do not include systematic effects whose importance
became evident only recently. The accuracy of these data is definitely lower
than the accuracy of the intermediate and high frequency data: neither
excludes nor confirms the existence of very large distortions, whose
amplitude, below 1 GHz, reach $\Delta T/T_{CMB}$ up to $\simeq 0.3 - 0.5$
\cite{sir2}.

The complete set of measured values of $T_{CMB}$ tells us that the matter
radiation mixture of our Universe was only perturbed by the energy injections
which occurred at $Z< 10^7$ . Future detections of large distortions at low
frequencies, whose existence has not yet excluded, should not modify this
picture. Analysis of the complete set of data (e.g.\cite{SaBu, NoSm}),
including the low frequency ones, shows in fact that, no matter if the energy
injection occurred at early or late times, the following upper limits hold:
\par\centerline{$\Delta E/E \leq 10^{-5}$ and $y \leq 10^{-5}$.}

\par\noindent
\begin{table}
\caption{Measurements of the CMB temperature below 5 GHz} \label{T1}
\vspace{5mm}
\begin{tabular}{llclllcl}
\hline
$\lambda~(cm)$ & $\nu~(GHz)$ & $T_{CMB}~ (K)$ & Ref.~~~~&$\lambda~(cm)$ & $\nu~(GHz)$ & $T_{CMB}~ (K)$ & Ref. \\
\hline 73.5~~ & 0.408~~ & ~3.7$\pm$1.2~ & \cite{HoSh67}&20.9 & 1.44 & 2.5$\pm$0.3 &\cite{PeSt69}\\
50.0 & 0.6 & 3.0$\pm$1.2 &\cite{Siro90}&20.7 & 1.45 & 2.8$\pm$0.6& \cite{HoSh66}\\
49.1 & 0.61 & 3.7$\pm$1.2 & \cite{HoSh67}&20.4&1.47&2.27$\pm$0.19&\cite{Bens}\\
47.2 & 0.635 & 3.0$\pm$0.5 & \cite{Stank}&15& 2.0 & 2.5$\pm$0.3&\cite{PeSt69}\\
36.6 & 0.82 & 2.7$\pm$1.6 &\cite{Siro91}&15 & 2.0 & 2.55$\pm$0.54 & \cite{Bers}\\
30.0 & 1.0 & 2.5$\pm$0.3 & \cite{PeSt69}&13.1 & 2.3 & 2.66$\pm$0.77& \cite{OtSt}\\
23.44&1.28&3.45$\pm$0.28& \cite{Subr00}&12 & 2.5 & 2.71$\pm$0.21 & \cite{Siro91}\\
21.26&1.41 & 2.11$\pm$0.38&\cite{Levi}&7.9 & 3.8 & 2.64$\pm$0.06 & \cite{Deam}\\
21.2 & 1.42 & 3.2$\pm$1.0 &\cite{PeWi67}&7.35& 4.08&3.5$\pm$1.0 &\cite{PeWi65}\\
21 &1.43& 2.65$\pm$0.32 & \cite{Stag96}&6.3 & 4.75 &2.79$\pm$0.07 & \cite{mand}\\
\hline
\end{tabular}
\vspace{5mm}
\end{table}

\section{TRIS experiment}

Because the low frequency data can still accommodate large distortions at
frequency and a positive detection at frequencies close to 1 GHz of a deep in
the distribution of $T_{CMB}$ versus $\nu$ would provide direct information on
the barionic density in our Universe, a few years ago our group set up TRIS,
a system of three absolute radiometers which measure $T_{sky}$, the absolute
temperature of the sky, at 0.6, 0.82 and 2.5 GHz. Fitting the values of the
sky temperature measured at three frequencies, at different points on the sky
to a model, we expect to be able to disentangle the foregrounds from the
background and extract the temperature of the CMB \cite{tris}.
 The antennae of the TRIS radiometers were corrugated horns,
geometrically scaled, with the same beam ($HPBW = 18^o\times 23^o$) at the
three frequencies. The horns could be aimed at different elevations along the
meridian, moreover, when absolute measurements were planned, we cooled the
waveguide section to liquid helium temperature $T_{LHe}$, reducing the
thermal noise produced at the system front end. The cryogenic bath used to
cool the waveguide housed also a dummy load which provided a stable reference
level $T_{ref}\simeq T_{LHe}$. A second dummy load at ambient temperature
provided a warm signal $T_{warm}$, used for calibrations. Details of the
system and its performance are given elsewhere \cite{tris}.

The three radiometers were installed at Campo Imperatore, a site at 2000 m
a.s.l. ($lat. = 42^o N$) near the Gran Sasso underground Laboratory, who
provided logistical support. At the beginning of the experiment the level of
radio interferences at Campo Imperatore was reasonably low, unfortunately the
situation became gradually worse, therefore we were gradually forced to
reduce the receiver bandwith and to move continously in frequency to overcome
the growing level of noise.
\par~\par
The measurements went on a few years and included:
\par~\par
a)absolute measurements of the sky temperature

To first approximation (for complete formulae see \cite{tris})
\begin{equation}
T_{sky}(\alpha,\delta) \simeq T_{ant} - T_{gro} - T_{atm}
\end{equation}
where $T_{ant}$, $T_{gro}$ and $T_{atm}$ are the {\it antenna temperature},
the ground contribution and the contribution of the atmosphere, respectively.
\begin{equation}
T_{ant}\simeq [T_{ref} + (S_{sky} - S_{ref})
\frac{T_{warm}-T_{ref}}{S_{warm}-S_{ref}} -  T_{amb}(1 -
e^{-\tau})]/e^{-\tau}~~~
\end{equation}
where $S_{sky}$, $S_{ref}$ and $S_{warm}$ are the signals produced
respectively by the sky, the cold dummy load and the warm dummy load, with
$T_{amb} \simeq T_{ref} \simeq T_{LHe}$~~(at Campo Imperatore elevation
$T_{LHe}4.0 K$)

Absolute measurements were made preferably at night time and repeated at
different times throughout the year to get data from different regions of the
sky.
\par~\par
b)drift scans. Letting the sky drift through the antenna beam aimed at a fixed
elevation along the meridian, we got profiles at constant declination of the
sky temperature versus the right ascension. To remove the sun contamination
data collected at daytime were eliminated and observation at the same
declination were repeated six months apart.

\par~\par
 c)measurements of the ground and atmospheric contributions.

The side and back lobes of the horns were minimized adding a band of
additional corrugations in the E-plane at the horn aperture. The resulting
beam was measured on a geometrically scaled model at 8.4 GHz. The ground
contribution was then evaluated convolving the measured beam with a blackbody
radiator at ambient temperature which filled all the directions below the
horizon profile (measured with a teodolite).

The atmospheric contribution was calculated combining models of the
atmospheric emission and profiles of atmospheric pressure, humidity and
temperature, measured daily by meteorological balloons \cite{aiel}. Results
are shown in Table 3.

\par\noindent
\begin{center}
\begin{table}
\caption{Values of the spectral index of the galactic diffuse radiation
measured at $\delta = 42 N$ between 0.61 and 0.82 GHz} \label{T2} \vspace{5mm}
\begin{tabular}{cclcclccl}
\hline
 ~~~~~~$\alpha$ & $\gamma$ & ~~~$\sigma ~~~~$ &$\alpha$ & $\gamma$ & $~~~\sigma ~~~~~$&$\alpha$
 & $\gamma$ & $~~~\sigma ~~~~$\\
\hline
~~~~~~$0^h$ & 2.379 & ~0.005~~~~ &$8^h$  &3.034 & ~0.003~~~~&$16^h$ &2.469 & ~0.002~~~~\\
~~~~~~$1^h$ & 2.515 & ~0.009~~~~ &$9^h$  &3.100 & ~0.004~~~~&$17^h$ &2.662 & ~0.002~~~~\\
~~~~~~$2^h$ &2.835  & ~0.009~~~~ &$10^h$ &3.231 & ~0.008~~~~&$18^h$ &2.693 & ~0.002~~~~\\
~~~~~~$3^h$ &2.540  & ~0.009~~~~ &$11^h$ &2.983 & ~0.011~~~~&$19^h$ &2.768 & ~0.002~~~~\\
~~~~~~$4^h$ &2.438  & ~0.005~~~~ &$12^h$ &2.834 & ~0.008~~~~&$20^h$ &2.883 & ~0.003~~~~\\
~~~~~~$5^h$ &2.618  & ~0.003~~~~ &$13^h$ &2.646 & ~0.006~~~~&$21^h$ &2.658 & ~0.002~~~~\\
~~~~~~$6^h$ &2.827  & ~0.002~~~~ &$14^h$ &2.393 & ~0.005~~~~&$22^h$ &2.649 & ~0.003~~~~\\
~~~~~~$7^h$ &2.973  & ~0.002~~~~ &$15^h$ &2.194 & ~0.003~~~~&$23^h$ &2.540 & ~0.003~~~~\\
\hline

\end{tabular}
\vspace{5mm}
\end{table}
\end{center}

\section{Extraction of $T_{CMB}$ : preliminary results}
The sky temperature is a combination

\begin{equation}
T_{sky}(\alpha,\delta,\nu) = T_{CMB} + T_{gal}(\alpha,\delta,\nu)
\left(\frac{\nu}{\nu_o}\right)^{-\gamma} +
T_{egs}(\nu)\left(\frac{\nu}{\nu_o}\right)^{-\beta}
\end{equation}
\noindent of the CMB signal and the signal produced by the galactic diffuse
emission and the blend of unresolved extragalactic sources.

$T_{gal}$ and $T_{egs}$ have power law frequency spectra. The spectral index
$\gamma$ of $T_{gal}$ can be obtained analyzing the TRIS drift scans made at
two frequencies by the T-T plot method \cite{turt,zann}. The value of
$T_{egs}$ can be obtained from data in literature (\cite{Siro90} and
references therein). Having $\gamma$ and $T_{egs}(\nu)$ we can then extract
$T_{CMB}(\nu)$ modelling the data collected at three frequencies, in
different directions and disentangling it from the galactic signal.

TRIS observations went to a halt in May 2001, when we were forced to move our
equipment to a different site, because the Campo Imperatore observing site was
closed. We are now carrying on data analysis. Only the reduction of the 0.6
GHz has been concluded, while the analysis of the 0.82 and 2.5 GHz data is
still going on. Results so far obtained are given in Table 2 and 3. At the
moment we can only give the temperature of $T_{CMB}$ at 0.6 GHz: being based
on data at one frequency the accuracy is still limited ($27\%$). We expect,
when the complete set of observations will be fully analyzed, to bring to
$10\%$ the accuracy on $T_{CMB}$ below 1 GHz.

\par\noindent
\begin{center}
\begin{table}
\caption{Measured values of the temperatures of CMB and foregrounds at
$r.a.=09^h57^m , dec.= 42^o26'$ - preliminary results} \label{T3} \vspace{5mm}
\begin{tabular}{cccc}
\hline
$~~~~~~~\nu (GHz)$ & 0.61 & 0.82 & 2.50 \\
\hline
~~~~~~~$T_{sky}(K)$ & 8.53$\pm$0.09  & under analysis & under analysis \\
~~~~~~~$T_{atm}(K)$ & 1.06$\pm$0.02 & 1.18$\pm$0.02 & 1.37$\pm$0.03 \\
~~~~~~~$T_{gnd}(K)$ & 0.07$\pm$0.05 & 0.07$\pm$0.05 & 0.07$\pm$0.05 \\
~~~~~~~$T_{egs}(K)$ & 0.81$\pm$0.14 & 0.34$\pm$0.07 & 0.016$\pm$0.003 \\
~~~~~~~$T_{gal}(K)$ & 4.74$\pm$0.81 & 2.06$\pm$0.37 & 0.087$\pm$0.024 \\
~~~~~~~$T_{CMB}(K)$ & 2.98$\pm$0.80 & under analysis & under analysis \\
\hline
\end{tabular}
\vspace{5mm}
\end{table}
\end{center}

\section{Conclusions}

The frequency region close to 1 GHz of the CMB frequency spectrum  is
potentialy very interesting. Here in fact we can expect distortions whose
measurement can provide an important cosmological parameter, the barion
density $\Omega_b$. Because the data in literature are insuffiucient to
decide about the existence of this distortion, we set up TRIS an experiment
with the aim of measuring the CMB spectrum between 0.6 and 2.5 GHz.
Observations have been completed but data reduction is still going on. We
expect to reach an accuracy of $\simeq 10\%$ sufficient to improve the
present upper limits on the amplitude and frequency  of, or hopefully detect,
the expected deep in the distribution of $T_{CMB}$ versus $\nu$.

\noindent \textbf{Acknowledgments} TRIS has been supported by the Milano
Universities, the Ministry of Research and the Natonal Council of Research.
The Gran Sasso underground laboratories provided logistical support at Campo
Imperatore. We are indebted to many people at the Laboratories for help in
carrying on the observations and maintaining the equipment.

\end{document}